\documentclass[prb,twocolumn,aps,showpacs,10pt,floatfix]{revtex4-1}

\usepackage[pdftex]{graphicx} 
\usepackage{epstopdf}
\usepackage{verbatim}
\usepackage{color}
\usepackage{subfigure}
\usepackage{tabularx}
\usepackage{amsfonts}
\usepackage{wasysym}
\usepackage{bm}

\newcommand{\be}{\begin{equation}}
\newcommand{\bea}{\begin{eqnarray}}
\newcommand{\ee}{\end{equation}}
\newcommand{\eea}{\end{eqnarray}}

\newcommand{\rb}{\right)}
\newcommand{\lb}{\left(}

\begin{document}
\title{Weyl magnons in breathing pyrochlore antiferromagnets}

\author{Fei-Ye Li$^{1,\ast}$}
\author{Yao-Dong Li$^{2,\ast}$}
\author{Yong Baek Kim${}^{4}$}
\author{Leon Balents${}^5$}
\author{Yue Yu$^{3,6,7}$}
\author{Gang Chen$^{3,6,7,\dag}$}
\affiliation{${}^1$Institute of Theoretical Physics,
Chinese Academy of Sciences, Beijing 100190, People's Republic of China}
\affiliation{${}^2$School of Computer Science, Fudan University,
Shanghai, 200433, People's Republic of China}
\affiliation{${}^3$State Key Laboratory of Surface Physics and Department of Physics,
Fudan University, Shanghai 200433, People's Republic of China}
\affiliation{${}^4$Department of Physics, University of Toronto,
Canadian Institute for Advanced Research, Quantum Materials Program,
Toronto, Ontario MSG1Z8, Canada and
School of Physics, Korea Institute for Advanced Study, Seoul 130-722, Korea}
\affiliation{${}^5$Kavli Institute for Theoretical Physics, Santa Barbara, California 93106, USA}
\affiliation{${}^6$Center for Field Theory and Particle Physics, Department of Physics,
 Fudan University, Shanghai 200433, People's Republic of China}
\affiliation{${}^7$Collaborative Innovation Center of Advanced Microstructures,
Nanjing, 210093, People's Republic of China}

\date{\today}
\begin{abstract}
Frustrated quantum magnets not only provide exotic ground states
and unusual magnetic structures, but also support unconventional excitations
in many cases. Using a physically relevant spin model for a breathing pyrochlore
lattice, we discuss the presence of topological linear band
crossings of magnons in antiferromagnets.  These are the
analogs of Weyl fermions in electronic systems, which we dub Weyl
magnons.  The bulk Weyl magnon implies the presence of chiral
magnon surface states forming arcs at finite energy.  We argue that
such antiferromagnets present a unique example in which Weyl points
can be manipulated in situ in the laboratory by applied
fields.  We discuss their appearance specifically in the breathing
pyrochlore lattice, and give some general discussion of conditions
to find Weyl magnons and how they may be probed experimentally.  Our
work may inspire a re-examination of the magnetic excitations in
many magnetically ordered systems.
\end{abstract}
\pacs{}

\maketitle

\noindent It is commonly thought that the spin ordering pattern of a magnetic
insulator uniquely specifies the state of the
system~\cite{Anderson1984}, and indeed the ground state of such
materials is usually well-described by a simple product state of
little fundamental interest.  However, in view of recent developments
in the study of topological properties of periodic media~\cite{RevModPhys.82.3045,RevModPhys.83.1057},
it is possible that even such a product-like ground state can support
topologically non-trivial excited state band structure.  Topological
properties of bands have been studied previously for electrons in
solids governed by Schr\"odinger's equation~\cite{RevModPhys.82.3045,RevModPhys.83.1057}, for
photons in dielectric superlattices governed by Maxwell's equations~\cite{PhysRevX.5.031011,LuLing},
for phonons governed by Newton's equations~\cite{PhysRevX.5.031011},
and even for fractionalized spinon excitation in spin liquids~\cite{PhysRevLett.114.157202,PhysRevB.93.085101}.
Here we apply these ideas to magnons governed by the equations for spin waves in an
ordered antiferromagnet.  We consider a concrete magnetic system,
namely, the Cr-based breathing pyrochlore, and explicitly demonstrate
that it supports Weyl magnon excitations with a linear band touching
in the spin wave spectrum of the magnetic ordered phase.  The Weyl
magnon is analogous to a Weyl fermion~\cite{Wan2011,Burkov2011,Hasan,Dinghong} in
electronic systems, but has bosonic rather than fermionic statistics,
similar to Weyl points in photonic systems~\cite{LuLing}.
In contrast to the other
three categories of systems, the band structure of magnons in
antiferromagnets is highly tunable in situ by application of
readily available magnetic fields, which is a consequence of the
spontaneous symmetry breaking of the antiferromagnet ground state and
the relatively low energy scale for magnetic interactions in most
solids.  Thus one can envision moving, creating, and annihilating Weyl
points in the laboratory in a single experiment.

\begin{figure}[ht]
\centering
    {\includegraphics[width=0.36\textwidth]{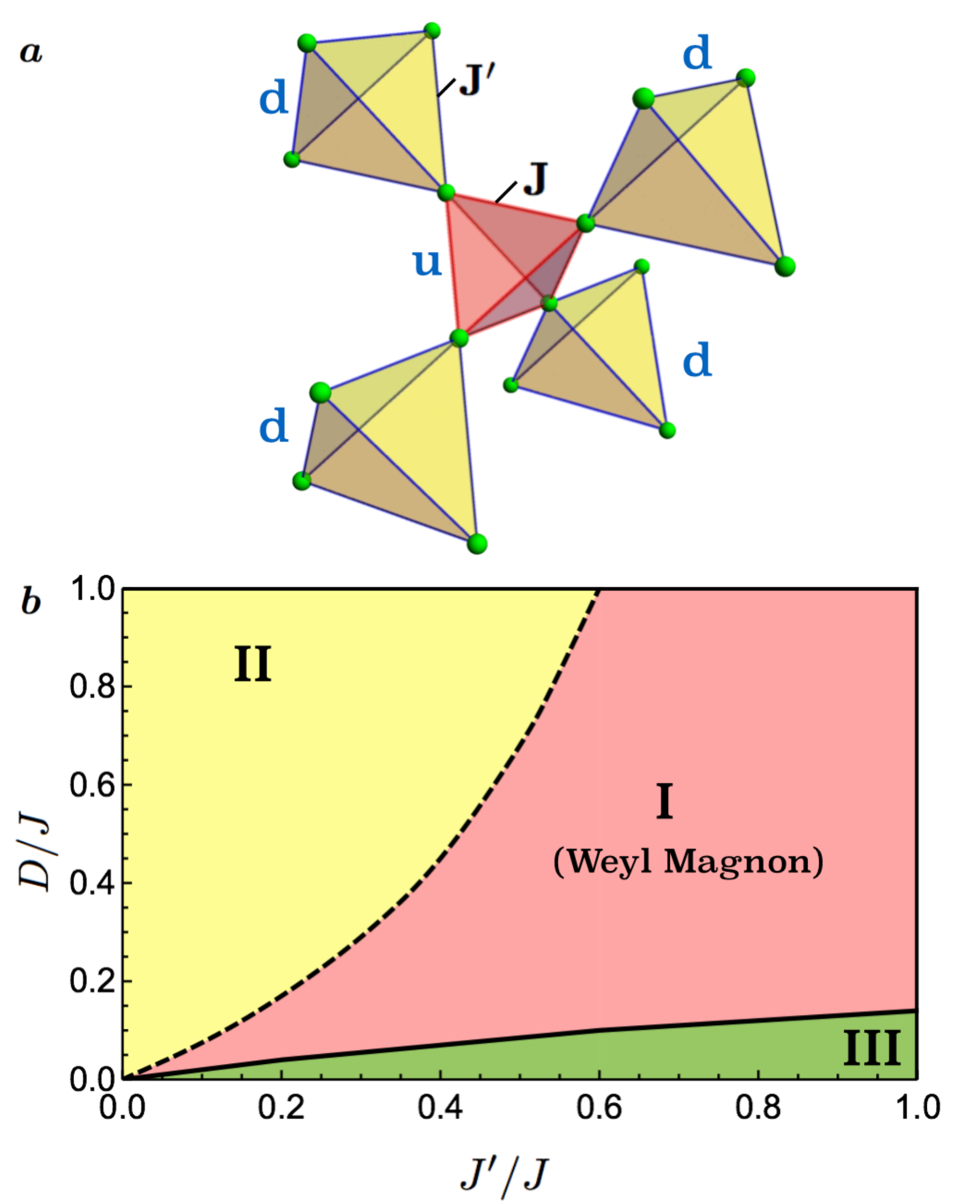}}
    \caption{{\bf The breathing pyrochlore and the phase diagram}.
	(a) The breathing pyrochlore. The letter u (d) refers 
	to the up-pointing (down-pointing) tetrahedra and $J$ ($J'$) 
	indicates the nearest-neighbour exchange couplings on the 
	up-pointing (down-pointing) tetrahedra.
	(b) The phase diagram. Region I and region II have the 
	same magnetic order and belong to the same phase, but the 
	magnetic excitations of the two regions are topologically 
	distinct. Region III has a different magnetic order. 
	The details of the phase diagram are discussed in the main text.
	}
\label{Fig1}
\end{figure}

To explore Weyl magnons, we focus on a concrete and physical model
system, the breathing pyrochlore antiferromagnet.
This is a generalization of the common pyrochlore
structure, which consists of a network of corner sharing tetrahedra,
with magnetic ions at the corners.  In the breathing pyrochlore,
alternate tetrahedra are uniformly expanded and contracted in
size~\cite{Yb_breathe,Cr_breathe2013,Cr_breathe2015,Savary2015ArXiv,JeffArXiv}.
As a result, the structure lacks an inversion center, and in general
up-pointing and down-pointing tetrahedral units are inequivalent.
We consider below a spin model for the breathing
pyrochlore, which generalizes and includes the uniform limit, and
displays Weyl points even in the uniform case.
We obtain the full phase diagram of this spin model and the magnetic excitations
in different phases. The experimental consequences of Weyl magnons and
the general conditions for their occurrence in spin systems
are predicted and discussed.
\\

\noindent{\bf Results}\\
\noindent{\bf{\small Spin model.}} We consider Cr$^{3+}$ ions in
the breathing pyrochlore lattice. There are several compounds with
this structure, including LiGaCr$_4$O$_8$ and LiInCr$_4$O$_8$, which
have been recently studied~\cite{Cr_breathe2013,Cr_breathe2015}.
In this $3d^3$ electron
configuration the orbital angular momentum is fully quenched and the
local moment is well described by the isotropic Heisenberg exchange
and a total spin $S=3/2$ according to Hund's rules.
The minimal spin model is given as
\bea H &=& J \sum_{\langle{ij}\rangle
  \in \text{u}} {\bf S}_i \cdot {\bf S}_j
+ J' \sum_{\langle{ij}\rangle \in \text{d}} {\bf S}_i \cdot {\bf S}_j \nonumber \\
&& + D \sum_{i} \lb {\bf S}_i \cdot {\bf \hat{z}}_i \rb^2, \label{eq1}
\eea
Since spin-orbit coupling is weak, the interaction
between the local moments is primarily
where we have supplemented the Heisenberg model with a local spin
anisotropy~\cite{PhysRevLett.109.016402} that is generically
allowed by the D$_{3d}$ point group
symmetry at the Cr site. The anisotropic direction $ {\bf \hat{z}}_i$ is the
local $[ 111 ]$ direction that points into the center of each
tetrahedron and is specified for each sublattice (Methods).  Here
$J$ and $J'$ are the exchange couplings between the nearest-neighbour
spins on the up-pointing and down-pointing tetrahedra (see Fig.~\ref{Fig1}),
respectively. The large and negative Curie-Weiss temperatures of the
Cr-based breathing pyrochlores indicate the strong AFM interactions,
hence we take $J > 0,J' > 0$. Because the up-pointing and
down-pointing tetrahedra have different sizes, one thus expects
$J \neq J'$.  In this work, however, we will study this model in a
general parameter setting. The AFM exchange interactions favor zero
total spin on each up-pointing (down-pointing) tetrahedron,
i.e. $\sum_{i \in \text{u}} {\bf S}_i = 0$
($\sum_{i \in \text{d}} {\bf S}_i = 0$). As for the regular pyrochlore
lattice~\cite{MoessnerChalker}, the classical ground state of the
exchange part of the Hamiltonian is extensively degenerate.
\\

\noindent{\bf{\small Ground states and quantum order by disorder.}}
We first consider easy-axis spin anisotropy with $D < 0$.
This favours the spin to be aligned with
its local $[ 111 ]$ axis. It turns out that this
condition can be satisfied while simultaneously optimizing the exchange interaction.
This gives a unique classical ground state
(up to a 2-fold degeneracy from the time reversal operation)
that has an all-in all-out magnetic order.
The magnetic excitation of this ordered state is fully gapped
and the energy gap ($\Delta$) is simply set by the
easy-axis spin anisotropy with $\Delta = 3|D|$ (Methods).

\begin{figure}
\includegraphics[width=0.5\textwidth]{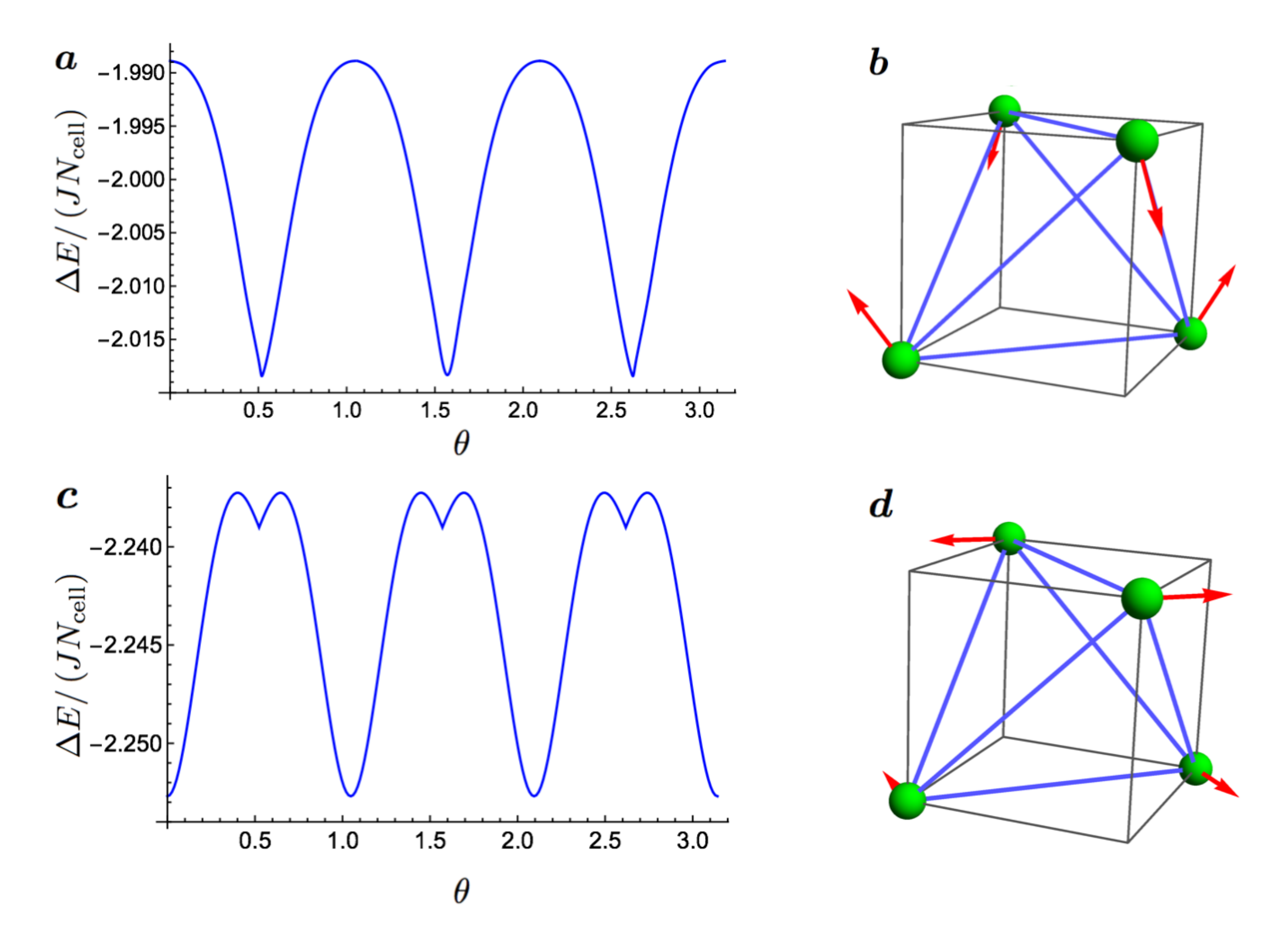}
\caption{{\bf Quantum zero point energy and the magnetic order}.
We have chosen the representative parameters in region I and region III
with $D = 0.2J$, $J' = 0.6J$ in (a) and $D=0.05J, J'=0.6J$ in (c),
respectively. (b) The magnetic order in region I and II with $\theta=\pi/2$
and the spins pointing along the local ${\bf \hat{y}}$.
(d) The magnetic order in region III with $\theta=0$ and
the spins pointing along the local $ {\bf \hat{x}}$.}
\label{Fig2}
\end{figure}

With the easy-plane anisotropy, $D>0$, the spin prefers to orient
in the $xy$ plane of the local coordinate system at each sublattice.
This requirement can also be satisfied while simultaneously optimizing
the exchange. Moreover, there exists an accidental $U(1)$ degeneracy
of the classical ground state that we parametrize as
\be
	{\bf S}^{\rm cl}_i \equiv S {\bf \hat{m}}_i
	= S(\cos \theta \, {\bf \hat{x}}_i + \sin \theta\, {\bf \hat{y}}_i),
	\label{eq2}
\ee
where ${\bf \hat{x}}_i$ (${\bf \hat{y}}_i$) is the unit vector along
the local $x$ ($y$) axis in the local coordinate system
at site $i$ (Methods), the unit vector ${\bf \hat{m}}_i$ points
in the local $xy$ plane, and the angular variable $\theta$
captures the $U(1)$ degeneracy.  This is the same form of degeneracy
found for the $S=1/2$ pyrochlore Er$_2$Ti$_2$O$_7$ in
Ref.~\onlinecite{Savary2012_disorder}, where it was noted that the degeneracy is
accidental, i.e. not protected by any symmetry, and hence will be
lifted by quantum fluctuations.
The same holds for the breathing pyrochlore, as we show now using
linear spin wave theory. We introduce the Holstein-Primakoff bosons
to express the spin operators as
${\bf S}_i \cdot {\bf \hat{m}}_i = S - a^\dagger_i a^{\phantom\dagger}_i$,
${\bf S}_i \cdot  {\bf \hat{z}}_i = (2S)^{1/2} (a^{\phantom\dagger}_i +
a^\dagger_i)/2$,
and
${\bf S}_i \cdot ({\bf \hat{m}}_i \times  {\bf \hat{z}}_i) = (2S)^{1/2}
(a^{\phantom\dagger}_i - a^\dagger_i)/(2i)$.
Keeping terms in the spin Hamiltonian $H$ up to the quadratic order in
the Holstein-Primakoff bosons, one can readily write down the spin
wave Hamiltonian as \bea H_{\rm sw} &= \sum_{\bf k} \sum_{\mu, \nu} [
&A_{\mu\nu} ({\bf k}) a^\dagger_{{\bf k}, \mu}
a^{\phantom\dagger}_{{\bf k}, \nu} + B_{\mu \nu}({\bf k})
a^{\phantom\dagger}_{-{\bf k}, \mu} a^{\phantom\dagger}_{{\bf k}, \nu} \nonumber \\
&&+ B^*_{\mu \nu}(-{\bf k}) a^\dagger_{{\bf k}, \mu} a^\dagger_{-{\bf
    k}, \nu} ] + E_{\rm cl},\label{eq3} \eea where $E_{\rm cl}$ is the
classical ground state energy, and $A_{\mu\nu}$, $B_{\mu\nu}$ satisfy
$A_{\mu\nu}({\bf k}) = A^*_{\nu\mu}({\bf k})$,
$B_{\mu\nu}({\bf k}) = B_{\nu\mu}(-{\bf k})$ and depend on the angular
variable $\theta$.  Although the classical energy $E_{\rm cl}$ is
independent of $\theta$ due to the $U(1)$ degeneracy, the quantum zero
point energy $\Delta E$ of the spin wave modes depends on $\theta$ and
is given by $\Delta E = \sum_{\bf k} \sum_\mu \frac{1}{2} 
[ \omega_\mu({\bf k}) - A_{\mu\mu}({\bf k})]$,
where $\omega_\mu({\bf k})$ is the excitation energy of the $\mu$-th
spin wave mode at momentum ${\bf k}$ and is determined for every
classical spin ground state. The minimum of $\Delta E$ occurs at
$\theta = \pi/6 + n\pi/3$ ($n \pi/3$) with $n \in \mathbb{Z}$ in
region I, II (region III). The discrete minima and the corresponding
magnetic orders are equivalent under space group symmetry
operations.  The $U(1)$ degeneracy of the classical ground states is
thus broken by quantum fluctuations. This is the well-known
phenomenon known as quantum order by
disorder~\cite{Savary2012_disorder,Zhitomirsky2014,Henley1987,Villain1980}.
The resulting optimal state is a non-collinear one in which each spin points
along its local $[ 112 ]$ ($[ 1\bar{1}0 ]$) lattice direction in 
region I, II (region III), see Fig.~\ref{Fig2}.

To obtain the phase diagram in Fig.~\ref{Fig1}, we have implemented the semiclassical
approach and included the quantum fluctuation within linear spin wave theory.
This treatment may underestimate the quantum fluctuation in the parameter regimes
when $J \gg J',D$ or $J' \gg J,D$. In the latter regimes, one may first consider
the tetrahedron with the strongest coupling and treat other couplings as
perturbations. The ground state in these regimes is likely to be non-magnetic
and will be addressed in the future work. For the purpose of the current work,
we will focus on the ordered ground states in Fig.~\ref{Fig1} and Fig.~\ref{Fig2}.
\\

\begin{figure}
\includegraphics[width=0.33\textwidth]{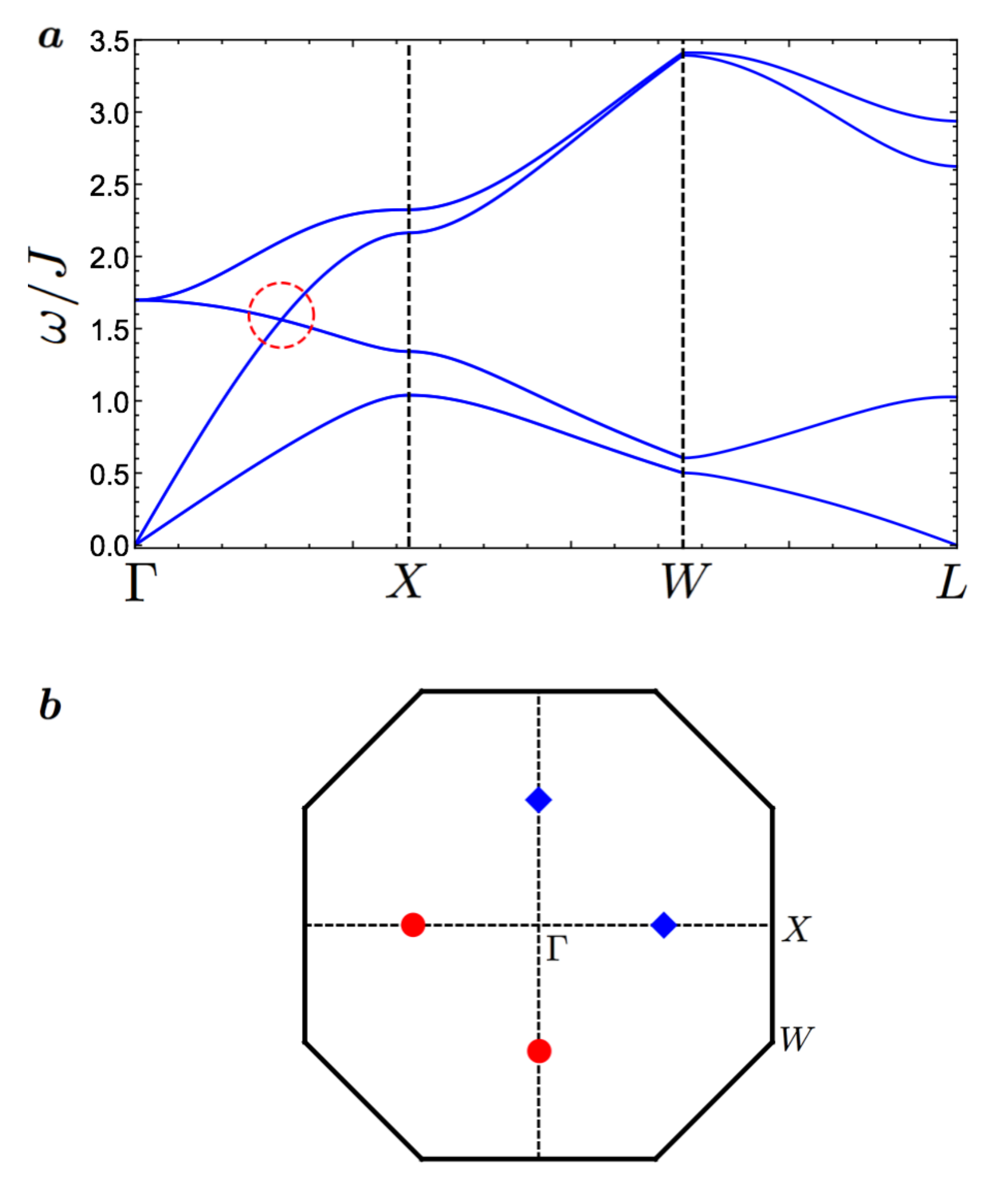}
\caption{{\bf The representative spin wave spectrum and the Weyl nodes of region I.}
(a) The spin wave spectrum along high symmetry momentum lines
with a linear band touching that is marked with a (red) dashed circle.
(b)  Four Weyl nodes are located at $[\pm k_0,0,0]$, $[0,\pm k_0,0]$ with $k_0=1.072\pi$
in the $xy$ plane of the Brillouin zone. The (red) circle has
an opposite chirality from the (blue) diamond.
In the figure, we have set $D = 0.2J$, $J' = 0.6J$ and $\theta=\pi/2$.
}
\label{Fig3}
\end{figure}

\noindent{\bf{\small Magnon Weyl nodes and surface states.}}
\noindent{Region} I and region II
have the same magnetically ordered structure with the same order parameter
and belong to the same phase.
Although the ground states are characterized by the same order parameter,
the magnetic excitations of the two regions are topologically distinct.
The magnetic excitation in region I has Weyl band touchings while the region II does not.
To further clarify this, we choose $\theta = \pi/2$
and thus fix the magnetic order to
orient along the ${\bf \hat{y}}$ directions of the local
coordinate systems. Using linear spin-wave theory,
we obtain the magnetic excitation spectrum with respect
to this magnetic state for region I and II.
In Fig.~\ref{Fig3}(a), we depict a representative excitation spectrum
along the high symmetry lines in the Brillouin zone for region I.

Two qualitative features are clear in the magnon spectrum of
Fig.~\ref{Fig3}(a).  First, we observe a gapless mode at the
$\Gamma$ point. This pseudo-Goldstone mode is an artifact of the
linear spin-wave approximation, and a small gap is expected to be
generated by anharmonic effects~\cite{Savary2012_disorder}. Secondly,
the spectrum in Fig.~\ref{Fig3}(a) has a linear band touching at a
point along the line between $\Gamma$ and X.  In fact, as we show in
Fig.~\ref{Fig3}(b), there are in total four such linear band
touchings.  The bands separate linearly in all directions away from
these touchings, which are thus Weyl nodes in the magnon
spectrum.  Just like Weyl nodes of non-degenerate electron
bands~\cite{Wan2011}, the magnon Weyl points are sources
and sinks of Berry curvature and are characterized by a discrete
chirality taking values $\pm 1$.  Unlike in an electronic Weyl semimetal,
where one can tune the Fermi energy to the Weyl nodes by
varying the electron density, the magnon Weyl nodes must necessarily
appear at finite energies because of the bosonic nature of magnons.

\begin{figure}
	\includegraphics[width=0.36\textwidth]{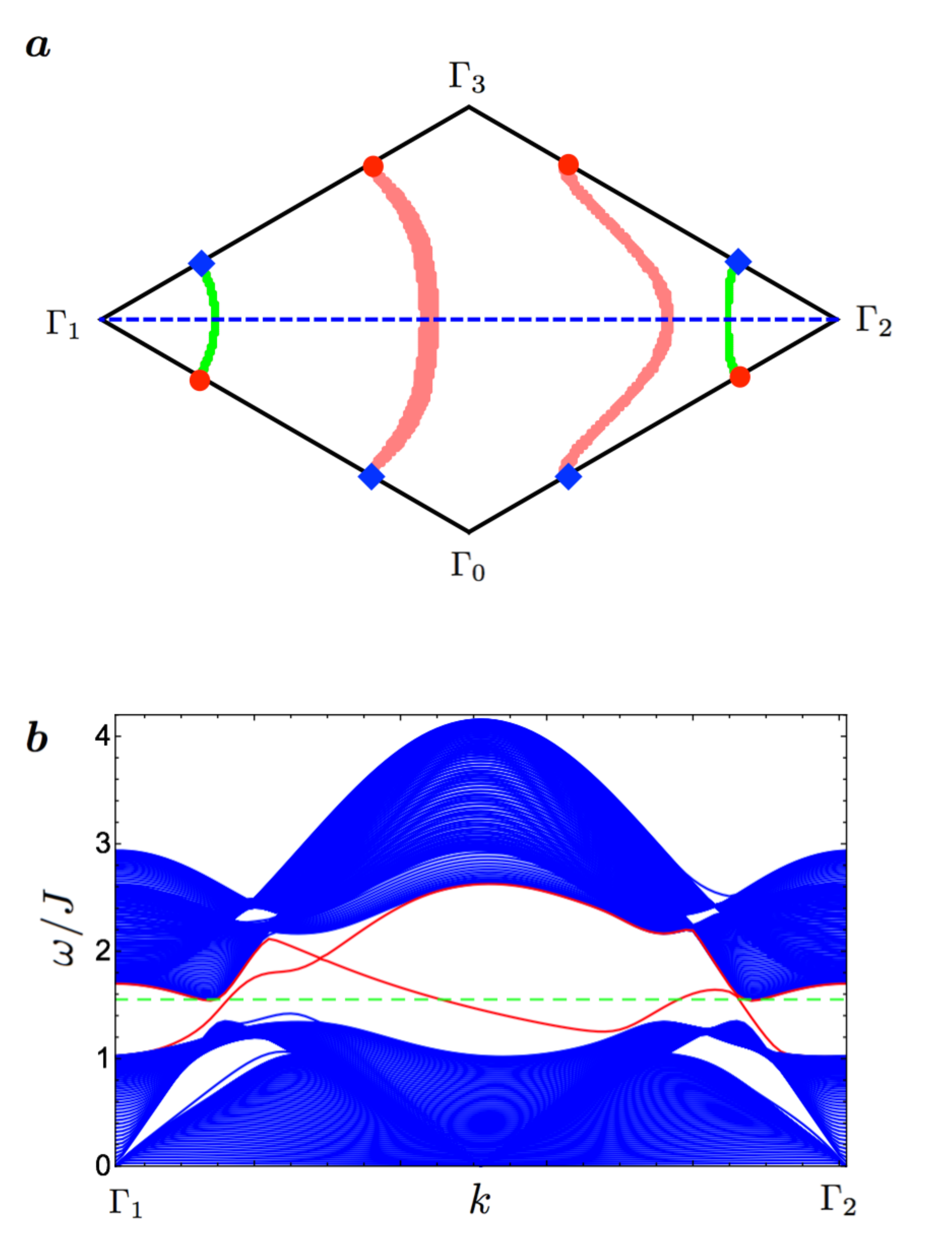}
	\caption{{\bf\small Surface states of a slab.}
	The slab is cleaved along $[11\bar{1}]$ surface, setting $D = 0.2J$, $J' = 0.6J$ and $\theta=\pi/2$.
	(a) Magnon arcs in the surface Brillouin zone. $\Gamma_0$ is the origin of the Brillouin zone
	 and two reciprocal lattice vectors are $\protect\overrightarrow{\Gamma_0\Gamma_1}=\frac{4\pi}{3}[2\bar{1}1]$,
	 $\protect\overrightarrow{\Gamma_0\Gamma_2}=\frac{4\pi}{3}[\bar{1}21]$.
	 The surface states with $E=E_{\text{Weyl}}$ form arcs connecting the Weyl nodes with different chiralities, where $E_{\text{Weyl}}$ is the energy of the bulk Weyl nodes.
	 States along the two pink longer (green shorter) arcs are localized in the top (bottom) surface.
	(b) The (blue) bulk magnon excitations and the (red) chiral surface states at $E_{\text{Weyl}}$
along $\protect\overrightarrow{\Gamma_1\Gamma_2}$.
	}
	\label{edge}
\end{figure}

Due to the bulk-edge correspondence, we expect magnon arc states bound
to any surface which possesses non-trivial projections of the bulk
Weyl points.   This is indeed observed in Fig.~\ref{edge}.  The chiral
magnon arcs appear at non-zero energy and connect the
bulk magnon Weyl nodes with opposite chiralities, as expected.

Once the magnon Weyl nodes emerge in the magnon spectrum, they are
topologically robust and exist over a finite regime in the parameter space.
We find that the magnon Weyl nodes exist in region I.
As the couplings are varied so that the boundary with
region II is approached, the magnon Weyl nodes move together and
annihilate in pairs when the boundary is reached.
In region II, there is no such (Weyl) band crossing,
qualitatively distinguishing region II from region I.
\\

\begin{figure}
	\includegraphics[width=0.5\textwidth]{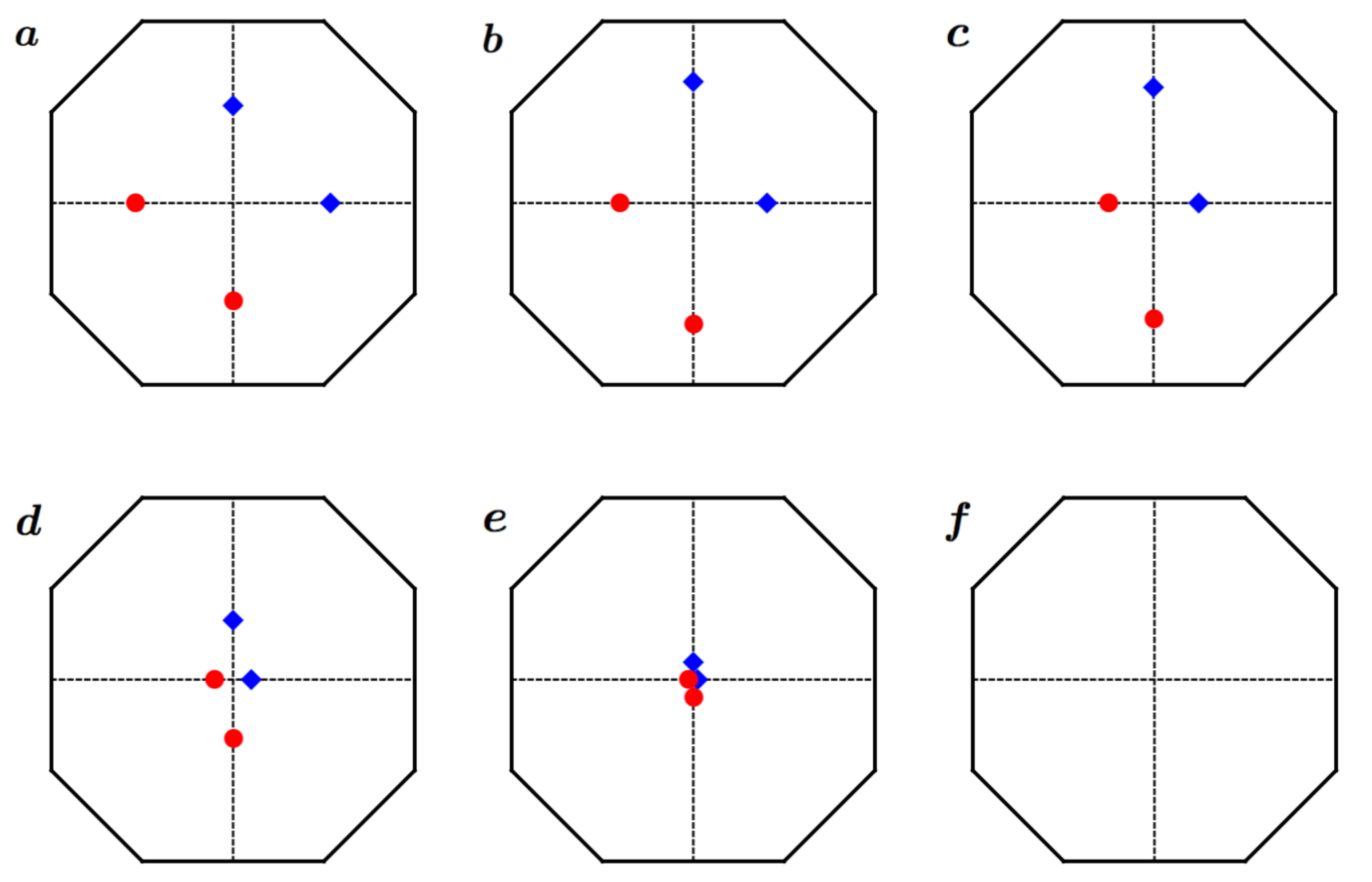}
	\caption
	{{\bf\small The evolution of Weyl nodes under the magnetic field}.
     Applying a magnetic field along the global $z$ direction,
     ${\bf B}=B {\bf \hat{z}}$, Weyl nodes are shifted but still in $k_z=0$ plane.
     They are annihilated at $\Gamma$ when magnetic field is strong enough.
     Red and blue indicate the opposite chirality.
     (a) to (f): $B=0$, $0.1J$, $0.5J$, $0.9J$, $1.0J$, $1.1J$.
     We have set $D = 0.2J$, $J' = 0.6J$ and $\theta=\pi/2$.
	}
	\label{magnet}
\end{figure}

\noindent{\bf{\small Manipulating Weyl nodes by external magnetic fields.}}
When we apply an external magnetic field to the system, the spin only
couples to the field via a Zeeman coupling.   This is quite different
from the case of electronic systems, in which a magnetic field also
has an orbital effect, which leads to cyclotron motion of electrons
and a transformation from ordinary bands into Landau ones.  In the
latter case, the meaning of quasi-momentum is irrevocably changed by
an applied field, and one cannot follow the Weyl point evolution with
field.  By contrast, since magnons are neutral, there is no orbital
effect, and quasi-momentum and the Weyl points themselves remain
well-defined even for strong fields.
Therefore, a magnetic field can be used to manipulate the Weyl nodes.
To demonstrate this explicitly, we focus on one specific classical order
in region I and apply a magnetic field along the global $z$ direction.
The magnetic field perturbs the classical ground state
and indirectly changes the spin-wave Hamiltonian.
As we show in Fig.~\ref{magnet}, the Weyl nodes are shifted gradually
and finally annihilated when the magnetic field is increased.
\\

\noindent{\emph{\bf Discussion}}\\
\noindent We have explicitly shown the presence of Weyl nodes in a simple and
physically relevant model for the breathing pyrochlore lattice
antiferromagnet.  Weyl points may also be present in other pyrochlores
for which the exchange is more complicated.  The spin wave spectra of
the highly anisotropic spin-1/2 pyrochlores Yb$_2$Ti$_2$O$_7$ and
Er$_2$Ti$_2$O$_7$ have been extensively studied~\cite{Ross11,Savary2012_disorder}.
Re-examined here in the light of topology, we see that they are present already in
the spin wave spectra of Yb$_2$Ti$_2$O$_7$ and
Er$_2$Ti$_2$O$_7$ in the external magnetic fields.
Thus we think that Weyl points can be present in
many magnetic materials of current interest.

Beyond these specific examples, we may ask what are the conditions
necessary to find Weyl points in the magnon spectrum?  In electronic
systems, these points are symmetry prevented, meaning that if both
inversion $\mathcal{P}$ and time-reversal symmetry $\mathcal{T}$ are
present, Weyl points cannot occur.  This is because in that case, a
two-fold Kramers' degeneracy of bands occurs, and any crossing must
involve two and not four bands.  For magnons, there is never a
Kramer's degeneracy.  This is because magnons are integer spin
excitations (even when the spin is not a good quantum number they are
superpositions of integer spin excitations), which do not obey
Kramer's theorem because $\mathcal{T}^2=+1$ in this case.  Moreover,
in general the magnetic order which underlies magnons already breaks
time-reversal symmetry.  This suggests that Weyl points may be
generically allowed.

However, there are some conditions under which Weyl points are
prohibited.  In particular, many magnetically ordered systems possess
not time-reversal but a complex conjugation symmetry $\mathcal{C}$.
This is the case for any Heisenberg Hamiltonian with a collinear
ordered ground state, but it can occur more generally. If, in
addition, the system possesses inversion symmetry $\mathcal{P}$, then
Weyl points are prohibited.  This can be understood from the Berry
curvature~\cite{PhysRevB.87.174427,PhysRevB.87.144101},
$\Omega_\mu({\bm k}) = i\epsilon_{\mu\nu\lambda} \langle \partial_\nu
{\bm k}|\partial_\lambda {\bm k}\rangle$, defined in terms of the
exact magnon eigenstates $|{\bm k}\rangle$ of a given magnon band.
The Berry curvature is an effective
magnetic field in momentum space, and a Weyl point is defined as a
delta-function source (divergence) of this curvature.  If
$\mathcal{P}$ is valid, one has $\Omega_\mu({\bm k}) =
\Omega_\mu(-{\bm k})$, while $\mathcal{C}$ implies $\Omega_\mu({\bm k}) =-
\Omega_\mu(-{\bm k})$.  Hence the combination requires
$\Omega_\mu({\bm k})=0$, prohibiting any Berry curvature at all, and
also obviously Weyl points.

This shows that in the simplest magnetically ordered systems, Weyl
points are not allowed.  There may be other conditions prohibiting
Weyl points, or constraining them.  A trivial condition is that one
needs at least two magnon bands to form Weyl points, which prohibits
them in some simple ferromagnets. In the case studied in this paper a two-fold
rotation axis locks the Weyl points along the $\Gamma$--X axes.
A full treatment of the necessary and sufficient conditions for Weyl
points may be part of a topological spin wave theory~\cite{151204902,Syzranov2015ArXiv},
to be developed in the future.

Now we turn to experimental implications.  The most natural probe of the
bulk magnon Weyl nodes as well as the surface magnon arc
states is inelastic neutron scattering. Because of the surface
dependence of the magnon arc states, one could study the
system with different slab geometries and surface orientations.
For example, for the [$11\bar{1}$] surfaces, one would observe
two disconnected arcs on both up and down surfaces (see Fig.~\ref{edge}).
In contrast, one would observe two loops across the
surface Brillouin zone for the [$110$] surfaces
because two pairs of Weyl nodes with different chiralities
are projected onto the same points (Methods).

The Weyl magnon can be potentially detected optically.  Close to the
Weyl nodes, a vertical transition can occur with arbitrarily small
energy.  Because the lower state is empty at zero temperature in
equilibrium, it may be beneficial to use a pump-probe approach to
measure the optical absorption. Then one may be able to observe
optical absorption at low frequency~\cite{PhysRevB.92.241108}, when
the lower magnon bands have enough population.  In addition to the
spectroscopic property, the presence of the Weyl magnon spectrum may
lead to a thermal Hall effect, just like the Weyl fermion that gives
rise to the anomalous Hall current in electronic
systems~\cite{YRan2011,Burkov2012}.  Furthermore, one could use
magnetic field to control thermal Hall signal~\cite{PhysRevB.89.134409,PhysRevLett.106.197202,
PhysRevB.90.024412}
despite the absence of
the Lorentz coupling of the spin to the external magnetic field.
Again due to population effects, the thermal Hall signal from Weyl
magnons will be suppressed at low temperature, but could be enhanced
by optical pumping.

Although the existing experiments suggest that both LiGaCr$_4$O$_8$
and LiInCr$_4$O$_8$ develop the antiferromagnetic long-range orders at
low temperature~\cite{Cr_breathe2013,Cr_breathe2015},
the precise structures of the magnetic order in these two systems
are not yet clear at this stage. Therefore, it is certainly of interest
to confirm the magnetic order and detect possible Weyl magnon
excitations in these systems and other three dimensional Mott insulators with
long range magnetic orders.

To summarize, we have studied a realistic spin model on the Cr-based
breathing pyrochlore lattice. We show that the combination of the
single-ion spin anisotropy and the superexchange interaction leads to
novel magnetically ordered ground states. Remarkably, the
magnetic excitations in a large parameter regime
develops magnon Weyl nodes in the magnon spectrum.
We expect that Weyl magnons may exist broadly in many
ordered magnets. We propose a number of experiments that can test
the presence of the Weyl magnons.\\

\noindent{\emph{\bf Methods}}\\
{\bf\small Local coordinate system.} The local coordinate system is defined for each
sublattice and is given in Table.~\ref{local_coord}.

\begin{table}[h]
 \begin{tabular}{c c c c }
   \hline
   \hline
	$\mu$ & $ {\bf \hat{x}}_\mu$ & $ {\bf \hat{y}}_\mu$ & $ {\bf \hat{z}}_\mu$  \\
	 1 & $\frac{1}{\sqrt{2}}[\bar{1}10]$ & $\frac{1}{\sqrt{6}}[\bar{1}\bar{1}2]$ & $\frac{1}{\sqrt{3}}[111]$\\
	 2 & $\frac{1}{\sqrt{2}}[\bar{1}\bar{1}0]$&$\frac{1}{\sqrt{6}}[\bar{1}1\bar{2}]$ & $\frac{1}{\sqrt{3}}[1\bar{1}\bar{1}]$ \\
	 3 & $\frac{1}{\sqrt{2}}[110]$ & $\frac{1}{\sqrt{6}}[1\bar{1}\bar{2}]$& $\frac{1}{\sqrt{3}}[\bar{1}1\bar{1}]$ \\
	 4 & $\frac{1}{\sqrt{2}}[1\bar{1}0]$ & $\frac{1}{\sqrt{6}}[112]$& $\frac{1}{\sqrt{3}}[\bar{1}\bar{1}1]$ \\
		\hline
		\hline
\end{tabular}
\caption[Table caption text]{{\bf\small The local axis
for the four sublattices of the breathing pyrochlore lattice.}
The letter $\mu$ refers to the sublattice, and
$({\bf \hat{x}}_\mu, {\bf \hat{y}}_\mu, {\bf \hat{z}}_\mu )$
defines the local coordinate system at the $\mu$-th sublattice.
}
\label{local_coord}
\end{table}

\begin{figure}[ht]
	\includegraphics[width=0.36\textwidth]{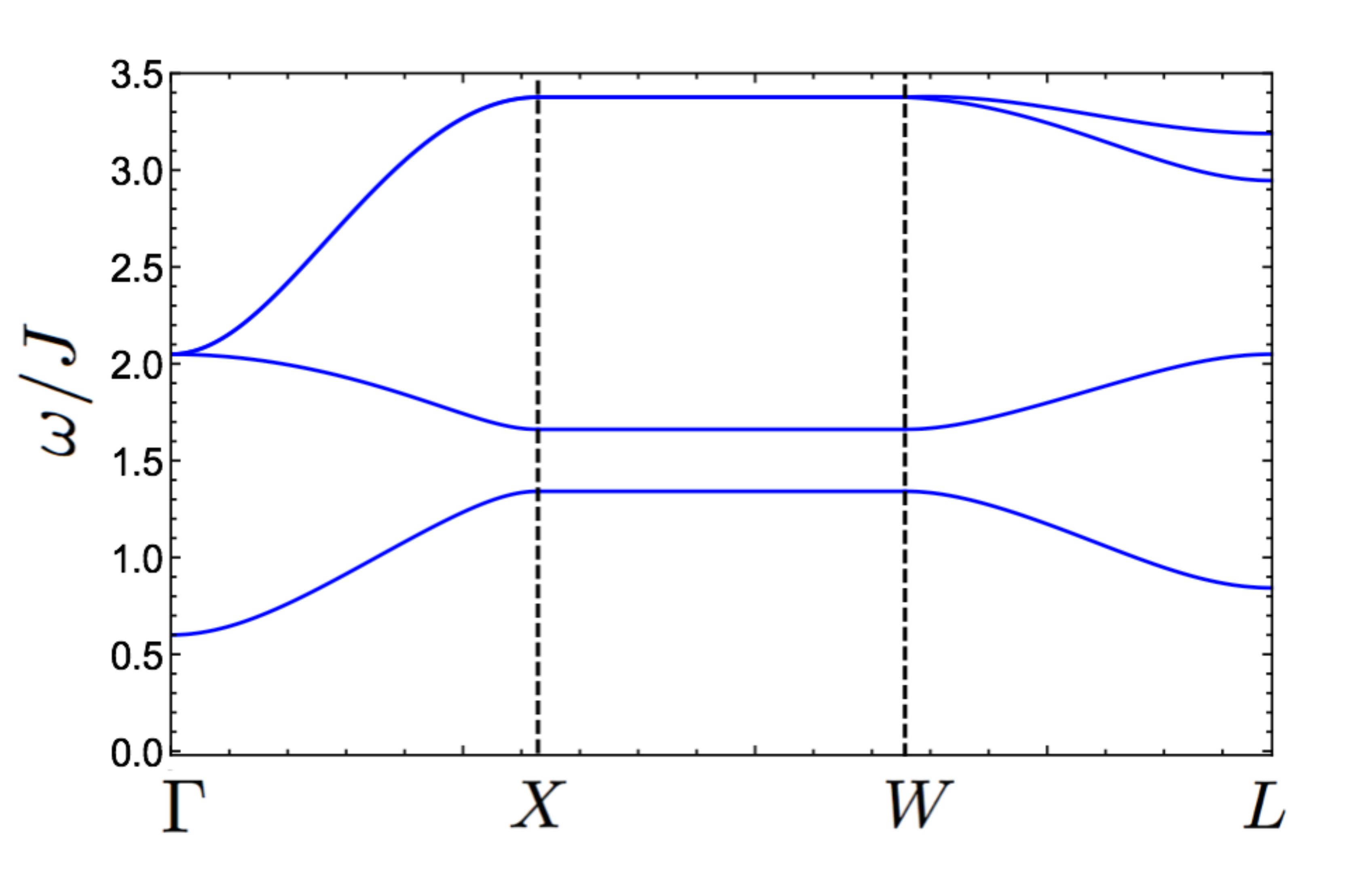}
	\caption
	{
     {\bf\small The spin wave spectrum of the all-in all-out state}.
     In this Figure, we set $D=-0.2J$, $J'=0.6J$.
	}
	\label{bandAIAO}
\end{figure}

\noindent{\bf\small Spin wave spectrum for the all-in all-out state.}
For the easy axis anisotropy with $D<0$, we have all-in all-out magnetic order,
and the spin wave $H_{\rm sw}$ in equation~(\ref{eq3}) is specified by the entries,
\begin{eqnarray}
A_{\mu \mu}({\bf k}) &=& S(-2D+J+J'), \\
A_{\mu \nu}({\bf k}) &=& -\frac{1}{3} S J_{\mu \nu} ,\\
B_{\mu \mu}({\bf k}) &=& 0,\\
B_{\mu \nu}({\bf k}) &=& \frac{1}{3} S J_{\mu \nu}  e^{i\phi_{\mu \nu}},
\end{eqnarray}
where $J_{\mu \nu}=J + J' e^{-i{\bf k}\cdot \left( {\bf b_{\nu}}-{\bf b_{\mu}} \right)}
\ (\mu \ne \nu)$,
${\bf b_1}=[ 000 ], {\bf b_2}=1/2[011 ],
{\bf b_3}=1/2[101 ],{\bf b_4}=1/2[ 110 ]$,
and
$\phi_{\mu \nu}=\phi_{\nu \mu} \ (\mu \ne \nu)$,
with
\be
\phi_{12}=\phi_{34}=-\frac{\pi}{3},
\phi_{13}=\phi_{24}=\frac{\pi}{3},
\phi_{14}=\phi_{23}=\pi.
\ee

The magnetic excitation of this ordered state is fully gapped
and the energy gap ($\Delta$) is simply set by the
easy-axis spin anisotropy with $\Delta = 3|D|$ (see Fig.~\ref{bandAIAO}).
\\

\begin{figure}[tp]
	\includegraphics[width=0.38\textwidth]{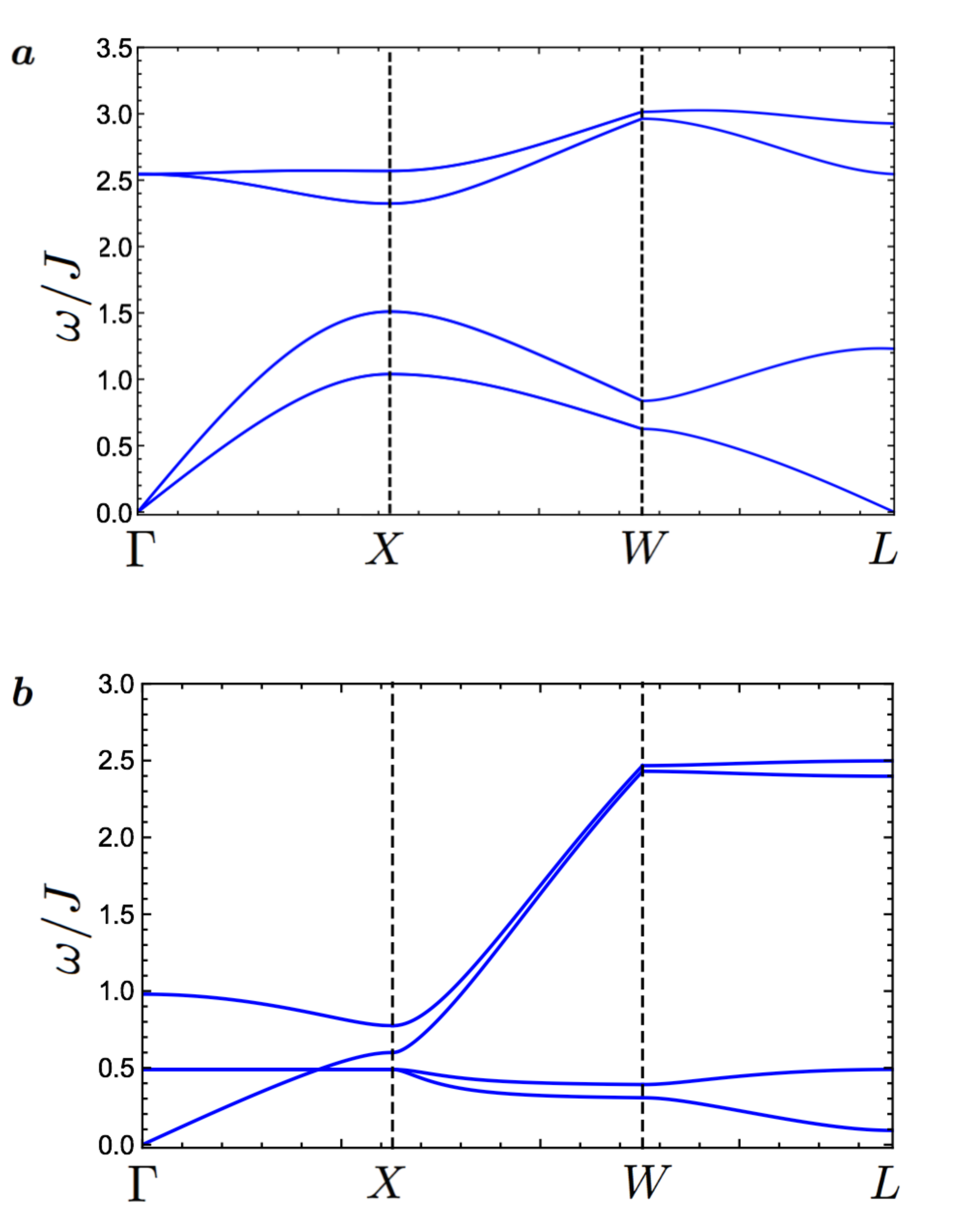}
	\caption
	{{\bf\small The spin wave spectrum of representative points in region II and region III}.
	In the figure, we have chosen the parameters as (a) $D=0.6J$, $J'=0.2J$, $\theta=\pi/2$,
	and	(b) $D=0.05J$, $J'=0.6J$, $\theta=\pi/3$.	}
	\label{band23}
\end{figure}

\begin{figure}[t]
	\includegraphics[width=0.30\textwidth]{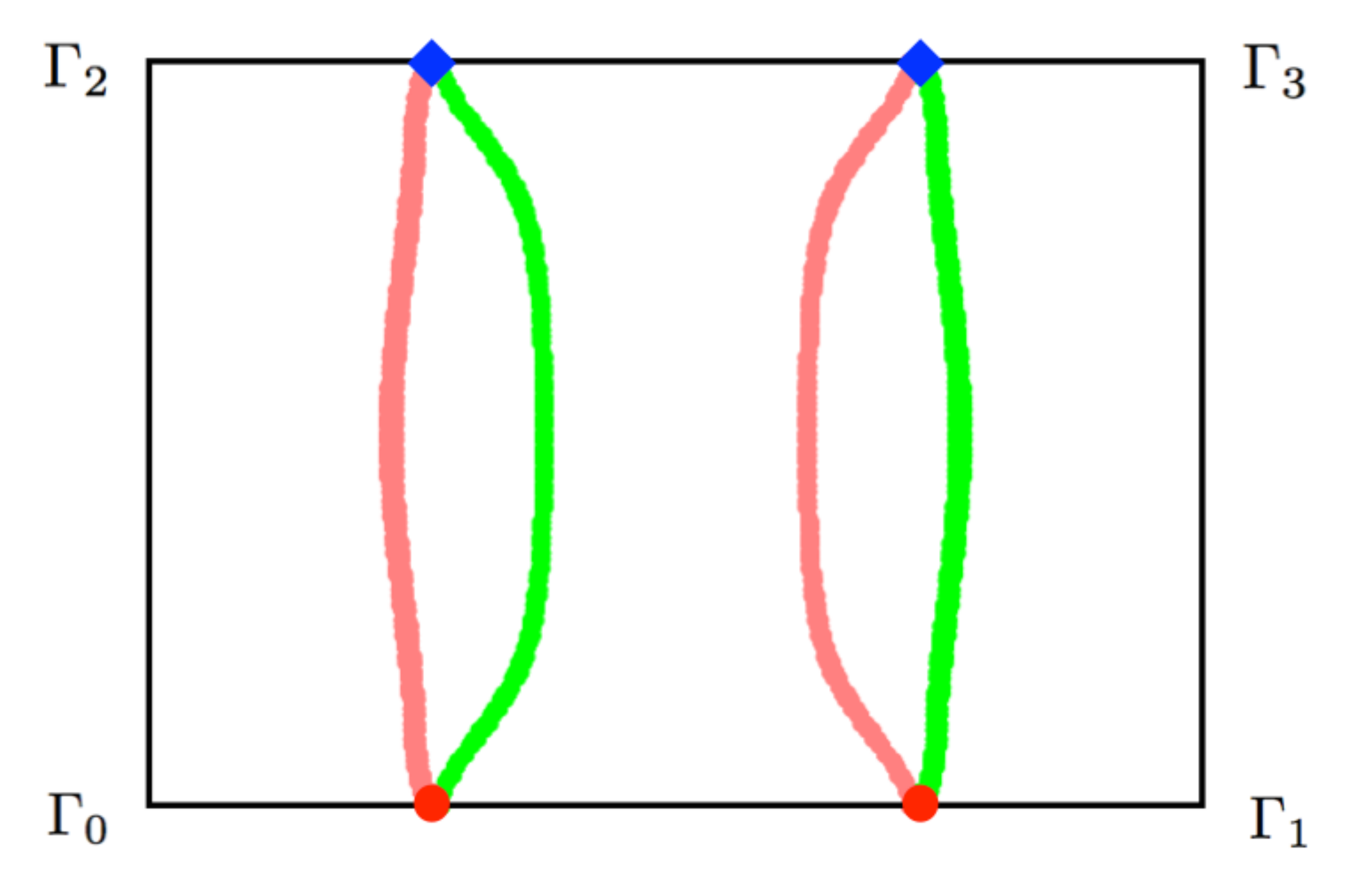}
	\caption{{\bf\small{Surface arc states of a slab.}}
		The slab is cleaved along $[110]$ surface, setting $D = 0.2J$, $J' = 0.6J$ and $\theta=\pi/2$. $\Gamma_0$ is the origin of the surface Brillouin zone, and two reciprocal lattice vectors are $\protect\overrightarrow{\Gamma_0\Gamma_1}=2\pi[1\bar{1}0]$, $\protect\overrightarrow{\Gamma_0\Gamma_2}=2\pi[001]$. The Surface states with $E=E_{\text{Weyl}}$ form arcs connecting the projections of Weyl nodes, where $E_{\text{Weyl}}$ is the energy of the bulk Weyl nodes. Note each pair of nodes are projected to the same position. Pink (Green) arcs are localized in one (the other) surface.}
	\label{fig8}
\end{figure}

\noindent{\bf\small In-plane ordered states.}
For the in-plane magnetic orders, the entries of
$H_{\rm sw}$ in equation~(\ref{eq3}) are given by
\begin{eqnarray}
A_{\mu \mu}({\bf k})&=&S(D+J+J'),\\
A_{\mu \nu}({\bf k})&=&-\frac{1}{3} S J_{\mu \nu}(1+\cos(2\theta+\phi_{\mu\nu})),\\
B_{\mu \mu}({\bf k})&=&\frac{1}{2}SD,\\
B_{\mu \nu}({\bf k})&=&\frac{1}{6} S J_{\mu \nu}
[\cos\left(2\theta+\phi_{\mu\nu}\right) \nonumber \\
 && -i2\sqrt{2}\cos\left(\theta-\phi_{\mu\nu}\right)
],
\end{eqnarray}
where $J_{\mu \nu}$ and $\phi_{\mu \nu}$ are the same as
the ones that are defined for the all-in all-out state.
In Fig.~\ref{band23}, we plot the spin wave spectrum for region II and region III.
For region III, there exists a band crossing between a dispersive band and
two (degenerate) flat bands from $\Gamma$ to X. This band crossing may
turn into Weyl band touchings if one includes extra spin interactions that
make the flat bands non-degenerate and dispersive.

Finally in Fig.~\ref{fig8}, we depict the surface arcs for the [110] surfaces in region II.
For this surface, each pair of nodes are projected to the same position, and the
surface arcs form two loops across the surface Brillouin zone and connect the Weyl nodes.

\noindent{\bf\small Data availability}\\
{ \noindent The data that support the findings of this study
are available from the corresponding author (G.C.) upon request.}

%


\vspace{1cm}

\noindent{\emph{\bf Acknowledgements}}\\
{\footnotesize \noindent{We} thank Xi Dai, Yang Qi,
R Shindou, Nanlin Wang, Zhong Wang, Xincheng Xie,
Fan Zhang, Fuchun Zhang and Yi Zhou for discussion and comments.
This work is supported by the Start-Up Fund of Fudan University
and the Thousand-Youth-Talent Program of People's Republic of China (FYL, YDL, GC),
the NSERC (YBK), the DOE Office of Basic Energy Sciences DE-FG02-08ER46524 (LB),
and the 973 Program of MOST of China 2012CB821402,
NNSF of China 11174298,11474061 (YY).}
\\

\noindent{\emph{\bf Contributions}}\\
{\footnotesize \noindent{G.C.} designed this project.
F.Y.L., Y.D.L. and G.C. performed the calculation.
F.Y.L., L.B. and G.C. wrote the manuscript.
All authors commented on the manuscript and the results.}\\

\noindent {\bf Additional information}\\
{\footnotesize \noindent {\bf Reprints and permissions}
information is available online at www.nature.com/reprints.}\\

{\footnotesize \noindent {F.Y.L.} and Y.D.L. contributed equally to this work.
Correspondence and requests for materials should be addressed to
G.C. (gangchen.physics@gmail.com).}\\

{\footnotesize \noindent \noindent {\bf Competing financial interests:}
The authors declare no competing financial interests.}


\begin{thebibliography}{35}%

 
    \bibitem{Anderson1984} Anderson, P. W. \textit{Basic Notions of Condensed Matter Physics} 2nd edn (Advanced Books Classics, Westview Press  / Addison-Wesley, 1997).

    \bibitem{RevModPhys.82.3045} Hasan, M. Z. \& Kane, C. L. \textit{Colloquium}\,:\,Topological insulators. \textit{Rev. Mod. Phys.} \textbf{82}, 3045--3067 (2010).

    \bibitem{RevModPhys.83.1057} Qi, X.-L. \& Zhang, S.-C. Topological insulators and superconductors. \textit{Rev. Mod. Phys.} \textbf{83}, 1056--1110 (2011).

    \bibitem{PhysRevX.5.031011} Peano, V., Brendel, C., Schmidt, M. \& Marquardt, F. Topological Phases of Sound and Light. \textit{Phys. Rev. X} \textbf{5}, 031011 (2015).

    \bibitem{LuLing} Lu, L. \textit{et al.} Experimental observation of Weyl points. \textit{Science} \textbf{349}, 622--624 (2015).

    \bibitem{PhysRevLett.114.157202} Hermanns, M., O'Brien, K. \& Trebst, S. Weyl Spin Liquids. \textit{Phys. Rev. Lett.} \textbf{114}, 157202 (2015).

    \bibitem{PhysRevB.93.085101} O'Brien, K., Hermanns, M. \& Trebst, S. Classification of gapless $\mathbb{Z}_{2}$ spin liquids in three-dimensional Kitaev models. \textit{Phys. Rev. B} \textbf{93}, 085101 (2016).

    \bibitem{Wan2011} Wan, X., Turner, A. M., Vishwanath, A. \& Savrasov, S. Y. Topological semimetal and Fermi-arc surface states in the electronic structure of pyrochlore iridates. \textit{Phys. Rev. B} \textbf{83}, 205101 (2011).

    \bibitem{Burkov2011} Burkov, A. A. \& Balents, L. Weyl Semimetal in a Topological Insulator Multilayer. \textit{Phys. Rev. Lett.} \textbf{107}, 127205 (2011).

    \bibitem{Hasan} Xu, S.-Y. \textit{et al.} Discovery of a Weyl fermion semimetal and topological Fermi arcs. \textit{Science} \textbf{349}, 613--617 (2015).

    \bibitem{Dinghong} Lv, B.-Q. \textit{et al.} Experimental Discovery of Weyl Semimetal TaAs. \textit{Phys. Rev. X} \textbf{5}, 031013 (2015).

    \bibitem{Yb_breathe} Kimura, K., Nakatsuji, S. \& Kimura, T. Experimental realization of a quantum breathing pyrochlore antiferromagnet. \textit{Phys. Rev. B} \textbf{90}, 060414 (2014).

    \bibitem{Cr_breathe2013} Okamoto, Y., Nilsen, G. J., Attfield, J. P. \& Hiroi, Z. Breathing Pyrochlore Lattice Realized in A-Site Ordered Spinel Oxides LiGaCr${}_4$O${}_8$ and LiInCr${}_4$O${}_8$. \textit{Phys. Rev. Lett.} \textbf{110}, 097203 (2013).

    \bibitem{Cr_breathe2015} Tanaka, Y., Yoshida, M., Takigawa, M., Okamoto, Y. \& Hiroi, Z. Novel Phase Transitions in the Breathing Pyrochlore Lattice: ${}^7$Li−NMR on LiInCr${}_4$O${}_8$ and LiGaCr${}_4$O${}_8$. \textit{Phys. Rev. Lett.} \textbf{113}, 227204 (2014).

    \bibitem{Savary2015ArXiv} Savary, L., Kee, H.-Y., Kim, Y. B. \& Chen, G. Magnetic phases and quantum spin ice on breathing pyrochlores. Preprint at http://arxiv.org/abs/1511.06972 (2015).

    \bibitem{JeffArXiv} Rau, J. G. \textit{et al.} Anisotropic exchange within decoupled tetrahedra in the quantum breathing pyrochlore Ba${}_3$Yb${}_2$Zn${}_5$O${}_{11}$. Preprint at http://arxiv.org/abs/1601.04104 (2016).

    \bibitem{PhysRevLett.109.016402} Chen, G., Hermele, M. \& Radzihovsky, L. Frustrated Quantum Critical Theory of Putative Spin-Liquid Phenomenology in 6H-B-Ba${}_3$NiSb${}_2$O${}_9$. \textit{Phys. Rev. Lett.} \textbf{109}, 016402 (2012).

    \bibitem{MoessnerChalker} Moessner, R. \& Chalker, J. T. Properties of a Classical Spin Liquid: The Heisenberg Pyrochlore Antiferromagnet. \textit{Phys. Rev. Lett.} \textbf{80}, 2929--2932 (1998).
 
    \bibitem{Savary2012_disorder} Savary, L., Ross, K. A., Gaulin, B. D., Ruff, J. P. C. \& Balents, L. Order by Quantum Disorder in Er${}_2$Ti${}_2$O${}_7$. \textit{Phys. Rev. Lett.} \textbf{109}, 167201 (2012).

    \bibitem{Zhitomirsky2014} Maryasin, V. S. \& Zhitomirsky, M. E. Order from structural disorder in the $XY$ pyrochlore antiferromagnet Er${}_2$Ti${}_2$O${}_7$. \textit{Phys. Rev. B} \textbf{90}, 094412 (2014).

    \bibitem{Henley1987} Henley, C. L. Ordering by disorder: Ground-state selection in FCC vector antiferromagnets. \textit{J. Appl. Phys.} \textbf{61}, 3962–-3964 (1987).

    \bibitem{Villain1980} Villain, J., Bidaux, R., Carton,  J. P. \& Conte, R. Order as an effect of disorder. \textit{J. Physique} \textbf{41}, 12631272 (1980).
 
    \bibitem{Ross11} Ross, K., Savary, L., Gaulin, B. \& Balents, L. Quantum Excitations in Quantum Spin Ice. \textit{Phys. Rev. X} \textbf{1}, 021002 (2011).
 
    \bibitem{PhysRevB.87.174427} Shindou, R., Matsumoto, R., Murakami, S. \& Ohe, J. Topological chiral magnonic edge mode in a magnonic crystal. \textit{Phys. Rev. B} \textbf{87}, 174427 (2013).
 
    \bibitem{PhysRevB.87.144101} Zhang, L., Ren, J., Wang, J.-S. \& Li, B. Topological magnon insulator in insulating ferromagnet. \textit{Phys. Rev. B} \textbf{87}, 144101 (2013).
 
    \bibitem{151204902} Fransson, J., Black-Schaffer, A. M. \& Balatsky, A. V. Magnon Dirac Materials. Preprint at http://arxiv.org/abs/1512.04902 (2015).
 
    \bibitem{Syzranov2015ArXiv} Syzranov, S. V., Wall, M. L., Zhu, B., Gurarie, V. \& Rey. A. M. Emergent Weyl quasi-particles in three-dimensional dipolar arrays. Preprint at http://arxiv.org/abs/1512.08723 (2015).

    \bibitem{PhysRevB.92.241108} Sushkov, A. B. \textit{et al.} Optical evidence for a Weyl semimetal state in pyrochlore Eu${}_2$Ir${}_2$O${}_7$. \textit{Phys. Rev. B} \textbf{92}, 241108 (2015).

    \bibitem{YRan2011} Yang, K.-Y., Lu, Y.-M. \& Ran, Y. Quantum Hall effects in a Weyl semimetal: Possible application in pyrochlore iridates. \textit{Phys. Rev. B} \textbf{84}, 075129 (2011).

    \bibitem{Burkov2012} Zyuzin, A. A., Wu, S. \& Burkov, A. A. Weyl semimetal with broken time reversal and inversion symmetries. \textit{Phys. Rev. B} \textbf{85}, 165110 (2012).

    \bibitem{PhysRevB.89.134409} Mook, A., Henk, J. \& Mertig, I. Magnon Hall effect and topology in kagome lattices: A theoretical investigation. \textit{Phys. Rev. B} \textbf{89}, 134409 (2014).

    \bibitem{PhysRevLett.106.197202} Matsumoto, R. \& Murakami, S. Theoretical Prediction of a Rotating Magnon Wave Packet in Ferromagnets. \textit{Phys. Rev. Lett.} \textbf{106}, 197202 (2011).

    \bibitem{PhysRevB.90.024412} Mook, A., Henk, J. \& Mertig, I. Edge states in topological magnon insulators. \textit{Phys. Rev. B} \textbf{90}, 024412 (2014).

\end{thebibliography}
\end{document}